\documentclass[journal,12pt,onecolumn,draftclsnofoot]{IEEEtran}
\usepackage[utf8]{inputenc}

\usepackage[utf8]{inputenc}
\usepackage{amsmath}
\usepackage{amsfonts}
\usepackage{amssymb}
\usepackage{algpseudocode}
\usepackage{algorithm}
\usepackage[pdftex]{graphicx}
\usepackage{epstopdf,epsfig}
\usepackage{enumerate}
\usepackage{pgfplots}
\usepackage{bbold}
\usepackage{subfigure}
\usepackage{mathtools}
\usepackage{amsthm}
\usepackage{wrapfig}
\usepackage{capt-of}
\usepackage[english]{babel}
\usepackage{url}
\usepackage{color}

\ifCLASSOPTIONcompsoc
  \usepackage[nocompress]{cite}
\else
  \usepackage{cite}
\fi

\theoremstyle{remark}

\newtheorem{theorem}{Theorem}[section]
\newtheorem{lemma}[theorem]{Lemma}
\newtheorem{remark}[theorem]{Remark}

\theoremstyle{definition}
\newtheorem{definition}{Definition}[section]

\newenvironment{Pf}{\textbf{Proof:}}{\hfill$\blacksquare$}

\newcommand{\Set}{\mathcal}
\title{Private Broadcasting: an Index Coding Approach}
\author{
\IEEEauthorblockN{Mohammed Karmoose, Linqi Song, Martina Cardone, Christina Fragouli}
University of California Los Angeles, Los Angeles, CA 90095 USA\\
Email: \{mkarmoose, songlinqi, martina.cardone, christina.fragouli\}@ucla.edu}

\begin{document}

\maketitle
\begin{abstract}
Using a broadcast channel to transmit clients' data requests may impose privacy risks. In this paper, we address such privacy concerns in the index coding framework. We show how a malicious client can infer some information about the requests and side information of other clients by learning the encoding matrix used by the server. We propose an information-theoretic metric to measure the level of privacy and show how encoding matrices can be designed to achieve specific privacy guarantees. We then consider a special scenario for which we design a transmission scheme and derive the achieved levels of privacy in closed-form. We also derive upper bounds and we compare them to the levels of privacy achieved by our scheme, highlighting that an inherent trade-off exists between protecting privacy of the request and of the side information of the clients.
\end{abstract}

\section{Introduction}

Consider a set of clients who share the same broadcast domain and wish to download data content from a server. 
Even though the content that they request may be publicly available, they wish to preserve the anonymity of their requests. 
For instance, assume that a client requests a video from YouTube related to a particular medical condition. 
If other clients learn about the identity of that request, this may then violate the privacy of that client. 
In this paper, we are interested in studying how to maintain the privacy of clients sharing a broadcast domain.


It is well established that coding across the content messages of the clients is needed to efficiently use the shared broadcast domain, as formalized in index coding~\cite{bar2011index}. 
A typical index coding instance consists of a server with $m$ messages, connected through a broadcast channel to a set of $n$ clients. 
Each client possesses a subset of the messages as side information and requires a specific new message.
The server then uses these side information sets to send coded transmissions, which efficiently deliver the required messages to the clients.

In this paper we claim that index coding poses a privacy challenge.
Consider, for example, that a server transmits $b_1+b_2$ to satisfy client~$1$. 
Since this is a broadcast transmission, other clients observing this transmission will infer that the request of client~$1$ is either $b_1$ or $b_2$, while the other message must belong to her side information.
This suggests that, although the clients can securely convey their requests to the server (e.g., through pairwise keys), a curious client may be able to infer information about the requests and/or side information sets of other clients by learning the encoding matrix used to generate the broadcast transmissions.


The first question we ask is: how much information does the encoding matrix in index coding reveal about the requests and the side information of other users? 
At a high level, one can think of the request and side information as two shared secrets between each client and the server, where one secret could be used to protect the other.
Therefore, as we also show in the paper, these two aspects exhibit a trade-off: maintaining a certain level of privacy on one aspect limits the amount of privacy level achieved on the other. 
We also ask: can we design index coding matrices that, for a given number of transmissions, achieve the highest possible level of privacy? 
How should these matrices be designed and how much privacy can they guarantee? 

In this paper, we take first steps in answering such questions. Our main contributions can be summarized as follows:
\\
\noindent 1) We propose an information-theoretic metric to characterize the levels of privacy that can be guaranteed. 
We then provide guidelines for designing encoding matrices and transmission strategies to achieve high privacy levels; \\
\noindent 2) We design an encoding matrix and characterize the maximum levels of privacy that it can achieve;  \\
\noindent 3) We derive universal upper bounds (i.e., which hold independently of the scheme that is used) on the maximum levels of privacy that can be attained;\\
\noindent 4) We consider a special case {of the problem and we characterize in closed-form the levels of privacy achieved by our scheme, which then we compare to the outer bounds, hence highlighting the privacy trade-off.}
%

\smallskip
\noindent\textbf{Related Work.}
In secure index coding~\cite{dau2012security}, the primal goal is to design strategies such that a passive external eavesdropper -- who wiretaps the communication from the server to the clients -- cannot learn any information about the messages.
Differently, in this work we seek to protect clients' \textit{privacy} against adversaries who wish to learn information about the identity of the requests and side information sets of the clients.

Recently, there has been a lot of effort trying to address privacy concerns in communication setups.
For instance, a set of relevant work has considered the problem of protecting privacy of a user against a database.
This problem was introduced in~\cite{chor1998private} and is known as {\it Private Information Retrieval} (PIR). Specifically,	
in PIR a client wishes to receive a specific message from a set of (possibly colluding) databases, without revealing the identity of the request. 
Towards this end, data request and/or storage schemes were designed~\cite{tajeddine2016private,freij2016private} and recently the PIR capacity was characterized~ \cite{sun2016capacity,banawan2016capacity}.

In cryptography, the {\it Oblivious Transfer} (OT) problem~\cite{Brassard1987} has a close connection to PIR~\cite{mishra2014oblivious}.
Specifically, in OT the goal is to protect both the privacy of the client against the server (i.e., as in PIR, the identity of the request of the client is not revealed to the server) and the privacy of the server against the client (i.e., the client learns only the requested message).
OT has also been used as a primitive to build techniques for secure multi-party computation~\cite{mishra2014oblivious}.

Different from these works, in this paper we seek to understand the privacy issues that can arise among clients who share the same broadcast domain. 
Specifically, we seek to design techniques that guarantee high levels of privacy both in the side information and in the request of a client against another curious client.
Given the different problem formulation, the techniques developed to solve the PIR and OT problems do not easily extend to our setup.



\smallskip
\noindent\textbf{Paper Organization.}
The paper is organized as follows.
In Section~\ref{sec::sys_model} we define our setup.
In Section~\ref{sec::def} we provide definitions and guidelines on how to design privacy-preserving transmission schemes and we derive fundamental upper bounds.
In Section~\ref{sec::encoding_matrix} we
present the design of a privacy-preserving matrix.
Based on this matrix, in Section~\ref{sec::achievable} we consider a specific scenario for which we propose a transmission scheme and assess its performance.
In Section~\ref{sec::conclusion} we conclude the paper.
Some of the proofs are delegated to the appendices.

\smallskip
\noindent\textbf{Notation.} 
Calligraphic letters indicate sets;
boldface lower case letters denote vectors and boldface upper case letters indicate matrices;
$|\Set{X}|$ is the cardinality of $\Set{X}$;
$[n]$ is the set of integers $\{1,\cdots,n\}$;
$2^{[n]}$ and ${[n] \choose s}$ are the power set and the set of all possible subsets of $[n]$ of size $s$, respectively;
for all $x \in \mathbb{R}$, the floor function is denoted with $\lfloor x \rfloor$;
for a sequence $X=\{X_1, \ldots, X_n\}$, $X_{\mathcal{S}}$ is the subsequence of $X$ where only the elements indexed by $\mathcal{S}$ are retained;
$\mathbf{0}_{i \times j}$ is the all-zero matrix of dimension $i \times j$;
$\mathbf{A}_{\mathcal{S}}$ is the submatrix  of $\mathbf{A}$ where only the columns indexed by $\mathcal{S}$ are retained;
$\text{span}(\mathbf{A})$ is the linear span of the columns of $\mathbf{A}$;
$H(X|y)$ is the entropy of the random variable $X$, conditioned on the \textit{specific} realization $y$;
${n \choose k} = 0$ if $k < 0$ or $k > n$;
logarithms are in base 2.

\section{Setup}
\label{sec::sys_model}
We consider a typical index coding instance, where a set of clients $\Set{N} = \{ c_{[n]} \}$, with $| \Set{N} | = n$, are connected to a server through a shared broadcast channel. 
The server has a database of messages $\Set{M} = \{b_{[m]}\}$, with $| \Set{M} | = m$.
Each client $c_i, i \in [n]$, is represented by a pair of random variables, namely: (i) $\bar{Q}_i \in [m]$ associated with the index of the message that $c_i$ wishes to download from the server and (ii) $\bar{{S}}_i \in {2^{[m]}}$, associated with the indices of the subset of messages she already has as side information.
We indicate with $\bar{q}_i$ and $\bar{\Set{S}}_i$ the realizations of $\bar{Q}_i$ and $\bar{S}_i$, respectively, which are chosen uniformly at random from their respective domains.
Clearly, $\bar{q}_i \notin \bar{\Set{S}}_i$.
We assume that the pairs ($\bar{Q}_i, \bar{S}_i$), $\forall i \in [n]$, are independent across $i \in [n]$.
%


\medskip

\noindent{\bf{Server Model.}}
We assume that the server knows the request and the side information of each client, i.e., it is aware of the realizations of the random variables $\bar{Q}_i = \bar{q}_i$ and $\bar{S}_i = \bar{\Set{S}}_i$, with $i \in [n]$.
Given this, the server seeks to satisfy
the requests of the clients through 
$T$ broadcast transmissions. 
The server employs linear encoding, i.e., each transmission consists of a linear combination of the $m$ messages, where the coefficients are chosen from a finite field $\mathbb{F}_L$ with $L$ being large enough.
This can be mathematically formulated as
$\mathbf{A} \mathbf{b} = \mathbf{y}$,
where $\mathbf{b} \in \mathbb{F}_{L}^m$ is the column vector of the $m$ messages, $\mathbf{A} \in \mathbb{F}_L^{T \times m}$ is the encoding matrix used by the server and $\mathbf{y} \in \mathbb{F}_L^T$ is the column vector with linear combinations of the messages.

Therefore, a {\it transmission scheme}
employed by the server 
consists of the following two components:
\begin{enumerate}[i)]
\item \textit{Transmission space}: a specific set $\mathcal{A}$ of encoding matrices designed to satisfy the clients and protect their privacy;
 \item \textit{Transmission strategy:} a function that, given ($\bar{q}_{[n]}, \bar{\Set{S}}_{[n]}$), determines the encoding matrix $\mathbf{A} \in \mathcal{A}$ to be used. 
We model the output of the function as a random variable $\mathbf{A}$ where $\mathbf{A} = \hat{\mathbf{A}}$ according to a probability distribution $p_{\mathbf{A}|\bar{Q}_{[n]},\bar{S}_{[n]}}(\hat{\mathbf{A}}|\bar{q}_{[n]},\bar{\Set{S}}_{[n]})$ that has to be designed. 
\end{enumerate}


\medskip

\noindent{\bf{Adversary Model.}}
We assume that some of the clients -- referred to as {\it eavesdroppers} -- are malicious.
Specifically, the eavesdroppers
are 
non-cooperative clients who, based on the broadcast transmissions they receive, are eager to 
infer information
about the requests and the side information sets of other clients.
Since the eavesdroppers do not cooperate, without loss of generality, we can assume that there is only one eavesdropper in the system, namely client $c_n$.
In addition, we assume that the eavesdropper $c_n$:
(i) is aware of 
both 
the transmission scheme employed by the server
and the underlying distribution based on which the clients obtain their requests and side information sets;
(ii) has infinite computational power;
(iii) knows the size of the side information set of each client, i.e., $s_i = |\bar{\Set{S}}_i|, i \in [n]$. 
This last assumption, which we make to simplify the analysis, provides pessimistic privacy guarantees with respect to a scenario where the eavesdropper does not have this information.
 

Based on this knowledge, the eavesdropper $c_n$ wishes to infer information about the request and side information of the other clients. 
Specifically, we denote with $Q_i$ and $S_i$ the random variables, which represent the eavesdropper's estimate of the request and side information of client $c_i$, respectively
and we let 
$p_{Q_i}(q_i)$ and $p_{S_i}(\Set{S}_i)$ be the corresponding probability density functions.
For ease of notation, in the rest of the paper, we drop the subscripts from the probability density functions while retaining the arguments.
Clearly, $Q_n=\bar{Q}_n$ and $S_n = \bar{S}_n$.
Before transmission, the eavesdropper is completely oblivious to $Q_i$ and $S_i$ for $i\in[n-1]$; we model this situation by having $p(q_i|s_i)$ and $p(\Set{S}_i|s_i)$ uniformly distributed over $[m]$ and ${[m] \choose s_i}$, respectively\footnote{In principle, in $p(q_i|s_i)$ and $p(\Set{S}_i|s_i)$ we should also have $q_n$, $\Set{S}_n$ and $s_{[n]\backslash \{i\}}$ in the conditioning. However, since ($\bar{Q}_i, \bar{S}_i$), $\forall i \in [n]$, are independent across $i$, we can safely drop this dependence.}. 
Then, by learning the specific encoding matrix $\hat{\mathbf{A}}$ employed by the server, the eavesdropper infers some information about the other clients, which is reflected in the conditional probability distributions $p(q_i|\hat{\mathbf{A}},s_{[n]},q_n,\Set{S}_n)$ and $p(\Set{S}_i|\hat{\mathbf{A}},s_{[n]},q_n,\Set{S}_n)$.
\medskip

\noindent{\bf{Privacy Metric.}}
We consider the amount of knowledge the eavesdropper has about the variables $Q_i$ and $S_i$ as a privacy metric. In particular, we evaluate
how far the uniform distribution is from the conditional distribution that the eavesdropper has after learning the encoding matrix $\hat{\mathbf{A}}$.
Let $X \in \{Q_{[n]}, S_{[n]}\}$. 
Then, inspired by the t-closeness metric for data privacy~\cite{li2007t}, we consider the \textit{Kullback–Leibler divergence} as a distance metric between the distributions $p(x|\hat{\mathbf{A}},s_i,q_n,\Set{S}_n)$ and $p(x|s_i)$, namely
\begin{align}
 &D_{\text{KL}}(p(x|\hat{\mathbf{A}},s_{[n]},q_n,\Set{S}_n) || p(x|s_i)) = \log(|\mathcal{X}|) - H(X |\hat{\mathbf{A}},s_{[n]},q_n,\Set{S}_n),
\end{align}
where $\mathcal{X}$ is the support of $X$ (note that the entropy used throughout the paper is conditioned on specific realizations). 
If $D_{\text{KL}}(p(x|\hat{\mathbf{A}},s_{[n]},q_n,\Set{S}_n) || p(x|s_i))  = 0$, i.e., $H(X | \hat{\mathbf{A}},s_{[n]},q_n,\Set{S}_n) = \log (|\mathcal{X}|)$), then the eavesdropper has no knowledge of the variable $X$.
Differently, larger values of $D_{\text{KL}}(p(x|\hat{\mathbf{A}},s_{[n]},q_n,\Set{S}_n) || p(x|s_i))$, i.e., smaller values of $H(X| \hat{\mathbf{A}}, s_{[n]},q_n,\Set{S}_n)$ indicate lower levels of privacy. 
Therefore, we consider $H(X| \hat{\mathbf{A}},s_{[n]},q_n,\Set{S}_n)$ as an indication of the level of privacy attained for the variable $X$. 
We focus on designing transmission schemes with guaranteed levels of privacy regarding three different quantities for each client:
\begin{enumerate}[i)]
 \item Privacy in the request, captured by $H(Q_i|\hat{\mathbf{A}}, s_{[n]},q_n,\Set{S}_n)$;
 \item Privacy in the side information, captured by $H(S_i|\hat{\mathbf{A}}, s_{[n]}, q_n, \Set{S}_n)$;
 \item Joint privacy, captured by $H(Q_i, S_i| \hat{\mathbf{A}}, s_{[n]}, q_n, \Set{S}_n)$.
\end{enumerate}
Therefore, our goal is to design a transmission scheme which provides privacy guarantees - in terms of the aforementioned metrics - for a given number of transmissions.

%

\section{Guidelines for Protecting Privacy}
\label{sec::def}

Based on the knowledge of ($\bar{Q}_{[n]}, \bar{S}_{[n]}$), the server chooses to use an encoding matrix $\mathbf{A} = \hat{\mathbf{A}}$ such that it satisfies all clients, i.e., it allows each client to decode her request using her side information set.
\begin{definition}
 A ($q$,$\Set{S}$) pair is said to be \textit{decodable in $\hat{\mathbf{A}}$} if, using $\hat{\mathbf{A}}$ as encoding matrix, message $b_q$ can be decoded knowing $b_{\mathcal{S}}$.
\end{definition}
\begin{definition}
 A $q$ (or $\Set{S}$) is said to be \textit{decodable in $\hat{\mathbf{A}}$} if there exists $\Set{S}$ (or $q$) such that ($q$,$\Set{S}$) is decodable in $\hat{\mathbf{A}}$.
\end{definition}
%
In order to design an encoding matrix that satisfies all clients, we rely on the following lemma -- a slight variation of~\cite[Lemma 4]{song2015content} -- which provides a decodability criterion for ($q$,$\Set{S}$) using a matrix $\hat{\mathbf{A}}$.
\begin{lemma}[Decodability Criterion]
\label{lemma::dec_criterion}
Let $\hat{\mathbf{A}}$ be the encoding matrix used by the server. 
Then, the pair ($q$,$\Set{S}$) is decodable in $\hat{\mathbf{A}}$ iff
${\hat{\mathbf{A}}}_{q} \notin \text{span}(\hat{\mathbf{A}}_{[m]\setminus \{ q \cup \Set{S} \}})$.
\end{lemma}
Lemma \ref{lemma::dec_criterion} provides a necessary and sufficient algebraic condition on whether a particular $(q,\Set{S})$ pair is decodable using a given encoding matrix. 
The eavesdropper, when trying to infer information about $c_i, i \in [n-1]$, can therefore apply this decodability criterion on all possible $(q_i,\Set{S}_i)$ pairs with $|\Set{S}_i|= s_i$, to determine the subset of pairs that are decodable using $\hat{\mathbf{A}}$.
In other words, since she knows that the request of client $c_i$ must be satisfied, then the actual $(\bar{q}_i,\bar{\Set{S}}_i)$ pair of client $c_i$ must belong to this set of decodable pairs. 
Thus, the size of the set of decodable pairs with side information sets of size $s_i$ determines the uncertainty that the eavesdropper has regarding the information of client $c_i$ and hence the attained levels of privacy for $c_i$.
Therefore, in order to maintain high levels of privacy, it is imperative to design encoding matrices with decodable sets of large sizes.

We next formalize this intuition. Towards this end, we
define the following three quantities: 
(i) $\Set{D}(\hat{\mathbf{A}},s_i)$, i.e., the set of decodable ($q_i,\Set{S}_i$) pairs in $\hat{\mathbf{A}}$ for client $c_i$;
(ii) $\Set{D}^Q(\hat{\mathbf{A}},s_i)$, i.e., the set of decodable $q_i$ in $\hat{\mathbf{A}}$ for client $c_i$, and 
(iii) $\Set{D}^S(\hat{\mathbf{A}},s_i)$, i.e., the set of decodable $\Set{S}_i$ in $\hat{\mathbf{A}}$ for client $c_i$. 
To better understand this notation, consider the following example.

\smallskip
\noindent\textbf{Example.} 
Consider $m = 5$, $n=2$ and $s_1 = 1$. 
If the server uses $\hat{\mathbf{A}}_1 = \left[\begin{matrix}1 & 0 & 0 & 0 & 0 \\ 0 & 0 & 1 & 0 & 0 \end{matrix} \right]$ as an encoding matrix, then $\Set{D}(\hat{\mathbf{A}}_1,1)=\left \{(1,i),(3,j) \right \}$ with $i \in[5]\backslash \{1\}$ and $j \in[5]\backslash \{3\}$, $\Set{D}^Q(\hat{\mathbf{A}}_1,1) = \{1,3\}$ and $\Set{D}^S(\hat{\mathbf{A}}_1,1) = [5]$.
Now, suppose that the server uses $\hat{\mathbf{A}}_2 = \left[\begin{matrix}1 & 1 & 0 & 0 & 0 \\ 0 & 0 & 1 & 1 & 0 \end{matrix} \right]$. 
Then, $\Set{D}(\hat{\mathbf{A}_2},1)=\{(1,2),(2,1),(3,4),(4,3) \}$, $\Set{D}^Q(\hat{\mathbf{A}}_2,1) = \Set{D}^S(\hat{\mathbf{A}}_2,1)= [4]$.
Clearly, $|\Set{D}(\hat{\mathbf{A}}_1,1)|>|\Set{D}(\hat{\mathbf{A}}_2,1)|$ and $|\Set{D}^S(\hat{\mathbf{A}}_1,1)|>|\Set{D}^S(\hat{\mathbf{A}}_2,1)|$, but $|\Set{D}^Q(\hat{\mathbf{A}}_1,1)|<|\Set{D}^Q(\hat{\mathbf{A}}_2,1)|$.

\smallskip

\noindent With this, we have the following remark that relates the privacy metrics to the sizes of the decodable sets (see Appendix \ref{app::thm_sym} for details).
\begin{remark}
\label{thm::uniformity}
When the eavesdropper observes the encoding matrix $\hat{\mathbf{A}}$, then for all $i \in [n-1]$ and $s_i \in [m-1]$, we have
\begin{subequations}
\label{eq:GenUB}
  \begin{align}
  &H(Q_{i},S_{i}|\hat{\mathbf{A}},s_{[n]},q_n,\Set{S}_n) \leq \log |\Set{D}(\hat{\mathbf{A}},s_i)|, \label{thm_prelim_joint} \\
  &H(Q_i|\hat{\mathbf{A}},s_{[n]},q_n,\Set{S}_n) \leq \log |\Set{D}^Q(\hat{\mathbf{A}},s_i)|, \label{thm_prelim_Q} \\
  &H(S_i|\hat{\mathbf{A}},s_{[n]},q_n,\Set{S}_n) \leq \log |\Set{D}^S(\hat{\mathbf{A}},s_i)|. \label{thm_prelim_S}
  \end{align} 
\end{subequations}
Moreover, these bounds are tight iff the corresponding probability distributions are uniform.
Namely: 
\begin{enumerate}[i)]
\item eq.\eqref{thm_prelim_joint} is tight iff $p(q_i,\Set{S}_i|\hat{\mathbf{A}},s_{[n]},q_n,\Set{S}_n)$ is uniform over $(q_i,\Set{S}_i) \in \Set{D}(\hat{\mathbf{A}},s_i)$;
\item eq.\eqref{thm_prelim_Q} is tight iff $p(q_i|\hat{\mathbf{A}},s_{[n]},q_n,\Set{S}_n)$ is uniform over $q_i \in \Set{D}^Q(\hat{\mathbf{A}},s_i)$;
\item eq.\eqref{thm_prelim_S} is tight iff $p(\Set{S}_i|\hat{\mathbf{A}},s_{[n]},q_n,\Set{S}_n)$ is uniform over $\Set{S}_i \in \Set{D}^S(\hat{\mathbf{A}},s_i)$.
\end{enumerate}
\end{remark}
%
%
Remark~\ref{thm::uniformity} implies that the sizes of the decodable sets give an upper bound on the corresponding levels of the privacy metrics.
Moreover, one can show that the conditions i) to iii) in Remark~\ref{thm::uniformity} hold -- and hence bounds \eqref{thm_prelim_joint} to \eqref{thm_prelim_S} are tight -- if $p(\hat{\mathbf{A}}|\bar{q}_{[n]},\bar{\Set{S}}_{[n]})$ in the transmission strategy (described in Section~\ref{sec::sys_model}) is properly designed.
For instance, using Bayes' rule, it can be shown -- see Appendix \ref{app::thm_sym} for the details -- that condition i) is satisfied iff 
\begin{align*}
\sum\limits_{q_{\mathcal{K}},\Set{S}_{\mathcal{K}} \in  \prod\limits_{j \in \mathcal{K}}  \Set{D}(\hat{\mathbf{A}},s_j)}  p(\hat{\mathbf{A}}|q_{[n]},\Set{S}_{[n]},s_{[n]}), \ \mathcal{K}=[n-1]\backslash \{i\}
  \end{align*}
is the same for all $(q_i,\Set{S}_i) \in \Set{D}(\hat{\mathbf{A}},s_i)$.

From Remark~\ref{thm::uniformity}, it follows that the design of
privacy-preserving transmission schemes 
consists of
two main steps: 
(i) designing encoding matrices with large decodable sets and (ii) using transmission strategies which satisfy uniformity conditions and hence achieve maximum levels of privacy.

Based on the result in Remark~\ref{thm::uniformity}, we now derive universal upper bounds (i.e., which hold independently of the encoding matrix that the server uses) on the decodable sets and hence on the levels of the privacy metrics. In particular, we have

\begin{lemma}
\label{thrm::upper_bounds}
 For any $\hat{\mathbf{A}} \in \mathbb{F}_L^{T \times m}$ and $s_i \in [m-1]$, we have
\begin{subequations}
 \begin{align}
&|\Set{D}(\hat{\mathbf{A}},s_i)| \leq T {m \choose s_i} \eqqcolon \mathsf{UB}_{Q,S}, \label{UB_QS_Joint}
\\& |\Set{D}^Q(\hat{\mathbf{A}},s_i)| \leq m  \eqqcolon  \mathsf{UB}_{Q}, \label{UB_Q}
\\& |\Set{D}^S(\hat{\mathbf{A}},s_i)| \leq  {m \choose s_i}  \eqqcolon  \mathsf{UB}_{S}. \label{UB_S}
 \end{align}
\end{subequations}
\end{lemma}
\begin{Pf}
The upper bounds in~\eqref{UB_Q} and~\eqref{UB_S} simply follow by noticing that the size of a decodable set is upper bounded by the size of the support of the corresponding random variable.
We next prove the bound in~\eqref{UB_QS_Joint}. 
For a given encoding matrix $\hat{\mathbf{A}} \in \mathbb{F}_L^{T \times m}$, one can write $\Set{D}(\hat{\mathbf{A}},s_i) = \sum_{\Set{S}_i \in {[m] \choose s_i}} \Set{N}(\hat{\mathbf{A}},\Set{S}_i)$, where $\Set{N}(\hat{\mathbf{A}},\Set{S}_i)$ is the set of requests $q_i \in \Set{D}^Q(\hat{\mathbf{A}},s_i)$ for which the pair $(q_i,\Set{S}_i)$ is decodable.
According to Lemma~\ref{lemma::dec_criterion}, for each $q_i \in \Set{N}(\hat{\mathbf{A}},\Set{S}_i)$, ${\hat{\mathbf{A}}}_{q_i}$ is not in the span of ${\hat{\mathbf{A}}}_{[m]\setminus {\Set{S}_i \cup q_i}}$.
It is therefore straightforward to show that the columns of ${\hat{\mathbf{A}}}_{\Set{N}({\hat{\mathbf{A}}},\Set{S}_i)}$ are linearly independent. 
Thus, $|\Set{N}(\hat{\mathbf{A}},\Set{S}_i) | \leq T$ and hence we have $ |\Set{D}(\hat{\mathbf{A}},s_i)| \leq T {m \choose s_i}$.
\end{Pf}


\section{Design of a Transmission Space}
\label{sec::encoding_matrix}

\begin{figure}
\centering
 \includegraphics[width=0.6\columnwidth]{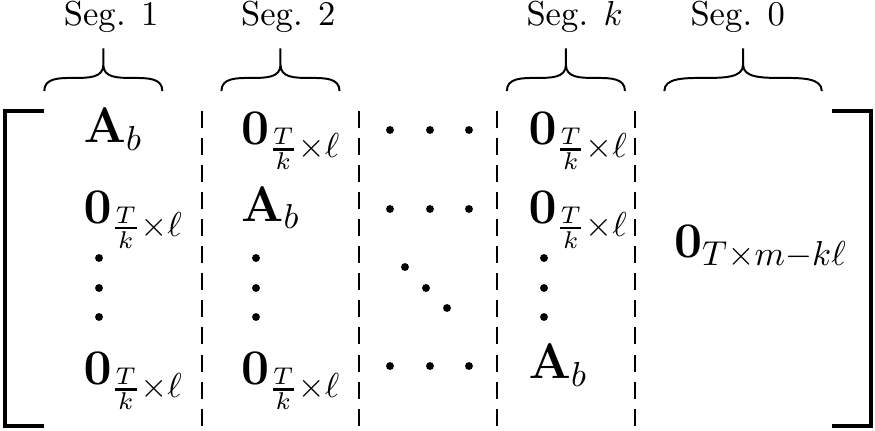}
 \caption{Design of the base matrix $\mathbf{A}^{\text{base}}$ for the achievable scheme.}
 \label{fig::base_matrix}
\end{figure}

In this section, we take first steps towards designing a privacy-preserving transmission scheme. Specifically, we design an encoding matrix, referred to as the {\it base} matrix $\mathbf{A}^{\text{base}}$.
Then, we populate the transmission space with the matrices obtained from $\mathbf{A}^{\text{base}}$ by taking all the permutations of its columns.
Our
%
%
%
design of $\mathbf{A}^{\text{base}}$ is based on the use of Maximum Distance Separable (MDS) codes. 
A generator matrix of an $[m,T]$ MDS code has the property that any $T \times T$ submatrix is full rank, i.e., any $T$ columns are linearly independent. Such matrices promise to provide large decodable sets. 
To see this notice that, for a given side information set $\Set{S}$ with $|\Set{S}| \geq m-T$, all requests in $[m]\setminus \Set{S}$ are decodable with $\Set{S}$. 
Therefore, if $\mathbf{B} \in \mathbb{F}_L^{T \times m}$ is a generator matrix of an $[m,T]$ MDS code, 
then, for all $s \geq m-T$, we have $|\Set{D}^Q(\mathbf{B},s)| = m$ and $|\Set{D}(\mathbf{B},s)| = m {m-1 \choose s} = O(m^s)$.
However, this scheme might require a prohibitively large number of transmissions $T$, especially when $m$ is large and $s$ is small compared to $m$.
To achieve high levels of privacy with $T$ that is not that large, we next propose the design of $\mathbf{A}^{\text{base}}$,
which is based on a block-MDS as shown in Figure~\ref{fig::base_matrix}
%
and structured as follows:

\begin{enumerate}[i)]
\item The columns of $\mathbf{A}^{\text{base}}$ are divided in $k+1$ segments, labeled as ``Seg. $0$ to $k$'', where $T$ is a multiple of $k$;
%
\item Segments from $1$ to $k$ consist of $\ell$ columns, where $\ell \leq \min\{ s_{\min} + T/k , \left\lfloor m/k \right\rfloor\}$, with $s_{\min} = \min_{i\in [n]} s_i$;
%
\item A matrix ${\mathbf{A}}_b \in \mathbb{F}_L^{\frac{T}{k} \times \ell}$ is constructed as the generator matrix of an $[\ell,T/k]$ MDS code; then, $\mathbf{A}_b$ is repeated $k$ times and positioned in $\mathbf{A}^{\text{base}}$ as shown in Figure~\ref{fig::base_matrix};
%
%
\item The rest of $\mathbf{A}^{\text{base}}$ is filled with zeros.
\end{enumerate}

Note that, for any number of clients $n$ and messages $m$, one can always find values of $k$, $\ell$ and $T$ so that $\mathbf{A}^{\text{base}}$ satisfies all clients (e.g., $k = 1, \ell = s_{\min}$ and $T = n$).

We now
analyze the performance of our proposed $\mathbf{A}^{\text{base}}$ in terms of the sizes of its decodable sets (see Appendix \ref{app::thm_dec_sets}).
These, by means of Remark~\ref{thm::uniformity}, provide
upper bounds on the levels of privacy that could be attained using $\mathbf{A}^{\text{base}}$.	

\begin{theorem}
\label{thm::thm_dec_sets}

 For $\mathbf{A}^{\text{base}}$ and any $s_i \in [m-1]$, we have
\begin{subequations}
 \begin{align}
  &H(Q_{i},S_{i}|\hat{\mathbf{A}},s_{[n]},q_n,\Set{S}_n) \leq \log \left( k \ell \!\!\!\! \sum\limits_{j = \ell - T/k}^{ \ell-1} \!\! {\ell-1 \choose j}{m - \ell \choose s_i - j} \right),  \\
  &H(Q_{i}|\hat{\mathbf{A}},s_{[n]},q_n,\Set{S}_n) \leq \left(k \ell\right). \label{scheme_dec_set} 
 \end{align}
\end{subequations}
\noindent where the bounds can be achieved by satisfying the uniformity conditions in Remark~\ref{thm::uniformity}
\end{theorem}

\begin{figure*}
 \centering
 \begin{minipage}{0.32\textwidth}
 \includegraphics[width=\columnwidth]{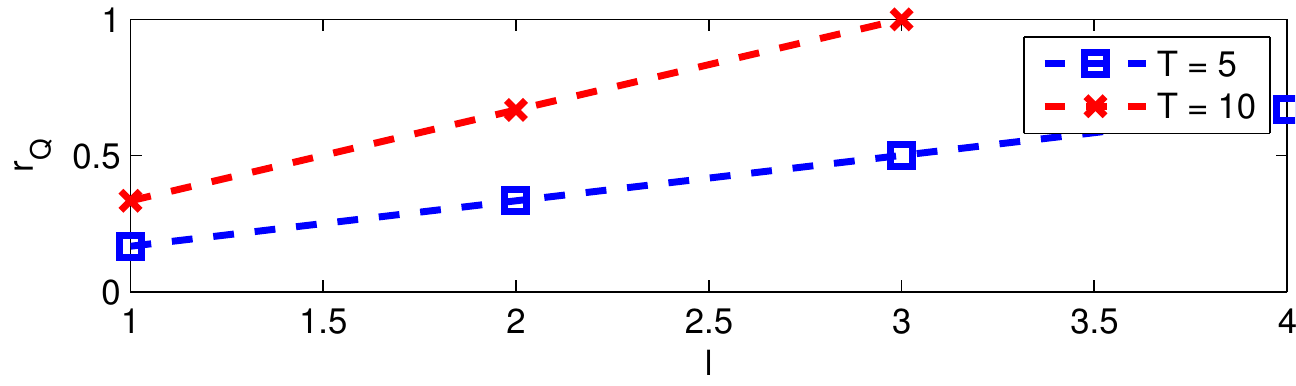}
 \end{minipage}
\begin{minipage}{0.32\textwidth}
 \includegraphics[width=\columnwidth]{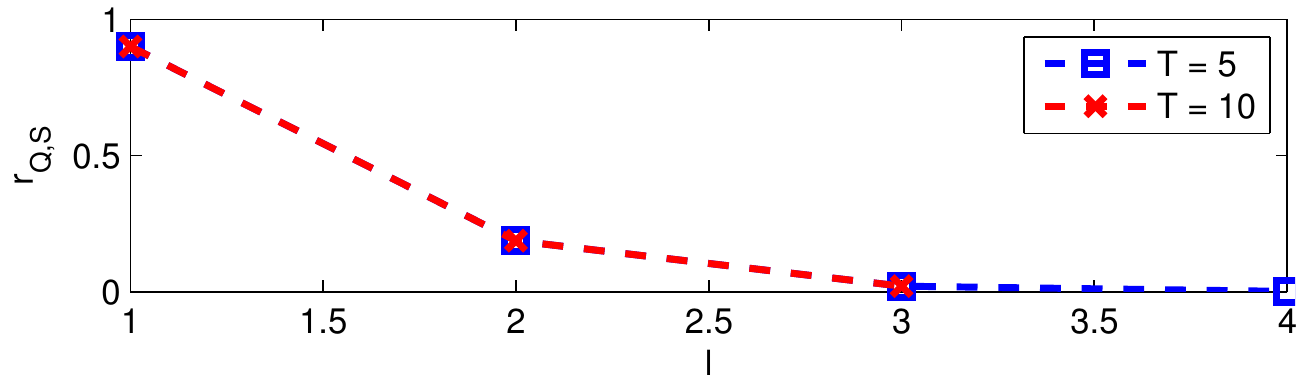}
 \end{minipage}
\begin{minipage}{0.32\textwidth}
 \includegraphics[width=\columnwidth]{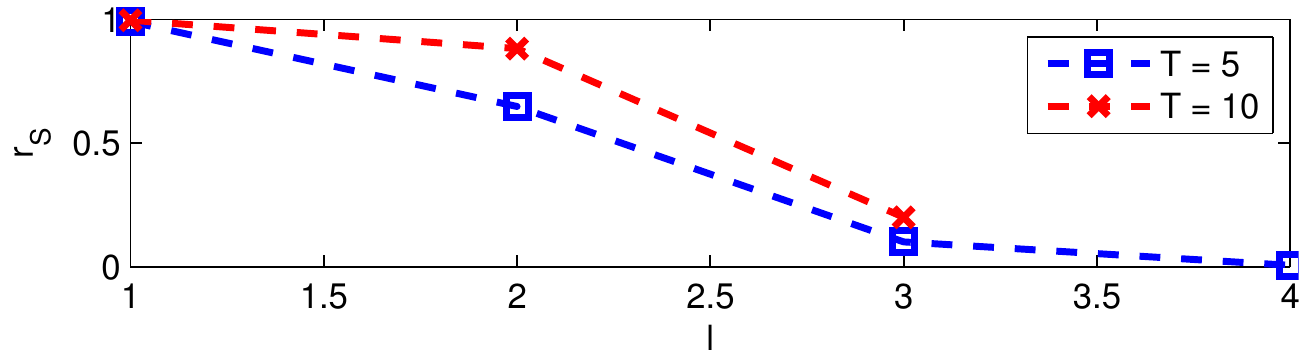}
 \end{minipage}
 \caption{Numerical evaluation of $\mathsf{r}_{Q}$, $\mathsf{r}_{S}$ and $\mathsf{r}_{Q,S}$ - $m = 30$ and $s = 3$.}
 \label{fig::eval}
\end{figure*}

In the next section we study a special scenario in which we use the transmission space here proposed (i.e., populated by the matrices obtained from $\mathbf{A}^{\text{base}}$ by taking all the permutations of its columns) and we design the transmission strategy.


\section{Transmission Strategy for a Special Case}
\label{sec::achievable}
In the previous section, we designed a transmission space that consists of all possible matrices obtained by permuting the columns of the matrix $\mathbf{A}^{\text{base}}$.
Thus, as discussed in Section~\ref{sec::sys_model}, in order to design a transmission scheme, we need to design a transmission strategy that selects which specific matrix to use according to a probability distribution.
However, designing such a transmission strategy that achieves the upper bounds in Remark~\ref{thm::uniformity} is non-trivial.
To get an analytical handle on the problem, we take a first step and consider a simplified model: we assume
$n = 2$ and an eavesdropper who does \textit{not} have a request.
Such a scenario can model a situation where the $n=2$ clients (the second of which is the eavesdropper) do not have a simultaneous request.

Since only one client needs to be satisfied, then we can use our proposed encoding matrix $\mathbf{A}^{\text{base}}$ with $k=T$ and $\ell \leq \min \{ s_1 + 1, \lfloor m/T \rfloor \}$, knowing that the client $c_1$ can always be satisfied by using the appropriate column-permutation of $\mathbf{A}^{\text{base}}$ (i.e., by ensuring that $\mathbf{A}^{\text{base}}_{q_1}$ is non-zero, and all other columns belonging to the same segment of $\mathbf{A}^{\text{base}}_{q_1}$ correspond to messages in $\Set{S}_1$).
In this case, $\mathbf{A}_b$ is a row vector of arbitrary non-zero values.
The following theorem (whose proof can be found in Appendix \ref{app::theorem_1}) then provides analytical guarantees on the attained performance of this scheme.

\begin{theorem}
\label{thm::achievable_scheme}
For the scheme described above, we have
\begin{subequations}
\label{main_thm}
 \begin{align}
 &H(Q_1,S_1|\hat{\mathbf{A}},s_1) =  \log T \ell { m - \ell \choose s_1 - \ell + 1} \eqqcolon \mathsf{LB}_{Q,S}, \label{eq:QSLB} \\
  & H(Q_1|\hat{\mathbf{A}},s_1) = \log T \ell \eqqcolon \mathsf{LB}_Q, \label{eq:QLB} \\
  & H(S_1|\hat{\mathbf{A}},s_1) =\log T \ell {{ m - \ell \choose s_1 - \ell + 1}} -  \mathsf{K} \eqqcolon \mathsf{LB}_S, \label{thm::HQ_HS_A11}\\
& \mathsf{K} = \sum\limits_{i=1}^T  \!\!{T-1 \choose i-1 }  \ell^{i-1} \frac{{m - i \ell \choose s_1 - i(\ell-1)}}{{ m - \ell \choose s_1 - \ell + 1}} \sum\limits_{x=1}^{i} (-1)^{i-x} {i \!-\! 1 \choose x \!-\! 1} \log x, \nonumber
 \end{align}
\end{subequations}
where $\hat{\mathbf{A}}$ is the column permutation of $\mathbf{A}^{\text{base}}$ that is used.
\end{theorem}
%
Note that the two quantities in~\eqref{eq:QSLB} and~\eqref{eq:QLB} meet the upper bounds that follow from Theorem~\ref{thm::thm_dec_sets} by applying the conditions in Remark~\ref{thm::uniformity}.
Moreover, in order to get the bounds in~\eqref{main_thm}, we used a transmission strategy for which $p(\hat{\mathbf{A}}|\bar{q}_1,\bar{\Set{S}}_1)$ is
uniform over all $\hat{\mathbf{A}}$ that satisfy $(\bar{q}_1,\bar{\Set{S}}_1)$ for all $(\bar{q}_1,\bar{\Set{S}}_1) \in \Set{D}(\hat{\mathbf{A}},s_1)$.
This is because, thanks to the special structure of $\mathbf{A}^{\text{base}}$, the number of column-permutations of $\mathbf{A}^{\text{base}}$ that satisfies a given $(\bar{q}_1,\bar{\Set{S}}_1)$ is equal for all $(\bar{q}_1,\bar{\Set{S}}_1)$.

We next analyze the performance of our scheme.
Towards this end, we define the following quantities:
\begin{itemize}
\item $\mathsf{G}_{Q,S} \coloneqq \log \left(\mathsf{UB}_{Q,S}\right) - \mathsf{LB}_{Q,S}$, $\mathsf{r}_{Q,S} = 2^{-\mathsf{G}_{Q,S}}$;
\item $\mathsf{G}_{Q} \coloneqq \log \left( \mathsf{UB}_{Q} \right) - \mathsf{LB}_{Q}$, $\mathsf{r}_{Q} = 2^{-\mathsf{G}_{Q}}$;
\item $\mathsf{G}_{S} \coloneqq \log \left( \mathsf{UB}_{S} \right) - \mathsf{LB}_{S}$, $\mathsf{r}_{S} = 2^{-\mathsf{G}_{S}}$.
\end{itemize}
Figure~\ref{fig::eval} shows an example of how the quantities $\mathsf{r}_{Q,S}$, $\mathsf{r}_{Q}$ and $\mathsf{r}_{S}$ behave as $\ell$ changes. 
Note that all these quantities are fractions {and hence the maximum level of privacy (y-axis) is~$1$.} 
{Figure~\ref{fig::eval} shows that as $\ell$ increases, higher values of privacy are attained in the requests (i.e., $\mathsf{r}_{Q}$ increases), but smaller levels of privacy are achieved in the side information (i.e., $\mathsf{r}_{S}$ decreases). 
This highlights a trade-off: maintaining a certain level of privacy on one aspect limits the amount of privacy level achieved on the other.} 
It is also noted that increasing $T$ increases the attained values of $\mathsf{r}_{Q}$ and $\mathsf{r}_{S}$ for the same value of $\ell$.
We believe that
the reason such increase does not occur in $\mathsf{r}_{Q,S}$ is because 
$\mathsf{UB}_{Q,S}$ in~\eqref{UB_QS_Joint}
is loose.

Next, we assess the performance of our scheme when the parameters of the system grow. 
%
%
%
We assume that $s_1 = c \cdot m$ and $\ell = b\cdot m + 1$, where $ b \leq c \leq \frac{m-1}{m}$. 
We consider two cases:
%
%

\noindent \textbf{Case I: $c = \frac{m-k_c}{m}$ where $k_c > 0$ is a constant.} 
In this case, full privacy in the request, side information and their joint can be achieved by using an $[m,T]$ MDS code with $T = k_c$.

\noindent \textbf{Case II: $c$ and $T$ are constants.}
In this case, by choosing $b = 0$, we get $\mathsf{G}_{Q} = \log \frac{m}{T} = O(\log m)$ and $\mathsf{G}_{Q,S} = \log \frac{{m \choose cm}}{T{ m-1 \choose cm}} = \log \frac{1}{T(1-c)} = O(1)$. 
Also, since conditioning reduces the entropy, we have $H(S_1|\hat{\mathbf{A}},s_1) \geq \text{eq.~\eqref{eq:QSLB}} - \text{eq.~\eqref{eq:QLB}}$,
which implies $\mathsf{G}_{S} \leq \log \frac{{m \choose cm}}{{ m-1 \choose cm}} = \log \frac{1}{1-c} = O(1)$. This
suggests that when $s_1$ grows as a constant fraction of $m$, then with a constant number of transmissions we can have almost perfect side information (and joint) privacy, but very little privacy in the request.
However, if we choose
$b = c$, then we get $\mathsf{G}_Q \leq \log \frac{1}{Tc} = O(1)$, $\mathsf{G}_{Q,S} =\mathsf{G}_{S}\leq \log {m \choose cm} = O(m \log m)$ since, under these conditions, $\mathsf{K}=0$ in~\eqref{thm::HQ_HS_A11}.
Thus, in this case almost full privacy is achieved in the request while very little privacy is attained in the side information (and in the joint).

%

\section{Conclusion}
\label{sec::conclusion}
We considered an index coding instance where some clients are malicious: they wish to learn information about the requests and side information of the other clients.
We showed how this privacy breach is possible by learning the encoding matrix used by the server. 
We proposed information-theoretic metrics to model the levels of privacy that can be guaranteed and
we designed an encoding matrix for protecting privacy.
Then, for a special case of the problem, we derived in closed-form the levels of privacy that our proposed scheme achieves.
We showed an inherent trade-off between protecting privacy of either the request or the side information set of the clients.

\bibliographystyle{IEEEtran}
\bibliography{Bib}

\appendices
\section{}
\label{app::thm_sym}
 We prove the result for the upper bound in \eqref{thm_prelim_joint}. Given $\hat{\mathbf{A}}$ and $s_i$, the set $\Set{D}(\hat{\mathbf{A}},s_i)$ consists of all possible ($q_i,\Set{S}_i$) pairs that could be the request/side information pair for $c_i$. Therefore, $p(q_i,\Set{S}_i|\hat{\mathbf{A}},s_{[n]},q_n,\Set{S}_n) = 0$ for all $(q_i,\Set{S}_i) \notin \Set{D}(\hat{\mathbf{A}},s_i)$. Therefore,
 \begin{align}
  H&(Q_{i},S_{i}|\hat{\mathbf{A}},s_{[n]},q_n,\Set{S}_n) = - \!\!\!\!\!\!\!\!\!\! \sum\limits_{(q_{i},\Set{S}_{i}) \in \Set{D}(\hat{\mathbf{A}},s_i)} \!\!\!\!\!\!\!\!\!\!  p(q_{i},\Set{S}_{i}|\hat{\mathbf{A}},s_{[n]},q_n,\Set{S}_n) \log p(q_{i},\Set{S}_{i}|\hat{\mathbf{A}},s_{[n]},q_n,\Set{S}_n) \leq \log | \Set{D}(\hat{\mathbf{A}},s_i)|  , \nonumber
 \end{align}
 \noindent thus proving \eqref{thm_prelim_joint}. Since $p(q_i,\Set{S}_i|\hat{\mathbf{A}},s_{[n]},q_n,\Set{S}_n) = 0$ for all $(q_i,\Set{S}_i) \notin \Set{D}(\hat{\mathbf{A}},s_i)$, then this upper bound is achieved if and only if $p(q_i,\Set{S}_i|\hat{\mathbf{A}},s_{[n]},q_n,\Set{S}_n)$ is uniform over for $(q_i,\Set{S}_i) \in \Set{D}(\hat{\mathbf{A}},s_i)$, thus proving the uniformity condition $i)$ on \eqref{thm_prelim_joint}. Similar arguments can be made to prove \eqref{thm_prelim_Q} and \eqref{thm_prelim_S}. 
 
 Next, we show that the uniformity conditions in $i)$-$iii)$ imply constraints on the design of the transmission strategy $p(\hat{\mathbf{A}}|q_{[n]},\Set{S}_{[n]})$.
  To see this, note that we can write

\begin{align}
 p(q_i,\Set{S}_i|&\hat{\mathbf{A}},s_{[n]},q_n,\Set{S}_n) = p(\hat{\mathbf{A}}|q_{\{i,n\}},\Set{S}_{\{i,n\}},s_{[n]}) \frac{p(q_i,\Set{S}_i|s_{[n]},q_n,\Set{S}_n)}{p(\hat{\mathbf{A}}|s_{[n]},q_n,\Set{S}_n)}, \nonumber
\end{align}
\noindent which follows by applying Bayes' rule. Since the probabilities in the fraction term do not depend on the value of $(q_i,\Set{S}_i)$ (note that $p(q_i,\Set{S}_i|s_{[n]})$ is uniform), then the uniformity condition $i)$ is satisfied if and only if the term $p(\hat{\mathbf{A}}|q_{\{i,n\}},\Set{S}_{\{i,n\}},s_{[n]})$ is the same for all $(q_i,\Set{S}_i) \in \Set{D}(\hat{\mathbf{A}},s_i)$. We can further write

\begin{align}
 p(\hat{\mathbf{A}}|q_{\{i,n\}},\Set{S}_{\{i,n\}},s_{[n]}) &= \!\!\!\!\!\!\!\!\!\!\! \sum\limits_{q_{\Set{K}},\Set{S}_{\Set{K}} \in  \prod\limits_{j \in \Set{K}} \Set{D}(\hat{\mathbf{A}},s_j)} \!\!\!\!\!\!\!\!\!\! p(\hat{\mathbf{A}}|q_{[n]},\Set{S}_{[n]},s_{[n]}) p(q_{\Set{K}},\Set{S}_{\Set{K}}|q_i,\Set{S}_i,s_{[n]}), \:\: \Set{K} = [n-1]\setminus {i}. \nonumber 
\end{align}

Note that the distribution $p(q_{\Set{K}},\Set{S}_{\Set{K}}|q_i,\Set{S}_i,s_{[n]})$ is assumed to be uniform and independent over $i \in [n]$. Therefore, to satisfy the uniformity condition, we must have the summation term on the Righ-Hand Side to be the same for all $(q_i,\Set{S}_i) \in \Set{D}(\hat{\mathbf{A}},s_i)$. This therefore imposes constraints on the transmission strategy used by the server. We can similarly show that the uniformity conditions on \eqref{thm_prelim_Q} and \eqref{thm_prelim_S} also impose constraints on the used transmission strategy.

 \section{}
 \label{app::thm_dec_sets}
 
 In order to prove Theorem \ref{thm::thm_dec_sets}, we need to characterize the quantities $|\Set{D}(\hat{\mathbf{A}},s)|$ and $|\Set{D}^Q(\hat{\mathbf{A}},s)|$, and therefore, using Remark~\ref{thm::uniformity} the result in Theorem~\ref{thm::thm_dec_sets} follows.
 
 \noindent\textbf{Characterizing $|\Set{D}^Q(\hat{\mathbf{A}},s)|$}:
 One can show that every request $q$ whose corresponding column $\mathbf{A}^{\text{base}}_q$ is non-zero has at least one side information set $\Set{S}$ with which $(q,\Set{S})$ is decodable in $\mathbf{A}^{\text{base}}$. If this in fact is true, then the result $|\Set{D}^Q(\mathbf{A}^{\text{base}},s)| = k\ell$ follows immediately, since we have $k\ell$ such requests.
 To prove this statement then, notice that $\ell \leq s_{\min} + T/k$. Then consider a side information set with $|\Set{S}| = s_{\min}$ and where all the elements of $\Set{S}$ correspond to columns of the same segment as $\mathbf{A}^{\text{base}}_q$. Therefore, the set of all columns of $\mathbf{A}^{\text{base}}$ belonging to the same segment as $\mathbf{A}^{\text{base}}_q$ and do not belong to $\Set{S}$ is of size $\ell - \Set{S} = T/k$. They are therefore linearly independent, and $q$ is decodable with $\Set{S}$.
 
  \noindent\textbf{Characterizing $|\Set{D}(\hat{\mathbf{A}},s)|$}:
 To prove the remaining quantity, notice that we can write $\Set{D}(\mathbf{A}^{\text{base}},s) = \sum_{q \in [m]} \Set{N}(\mathbf{A}^{\text{base}},q)$, where $\Set{N}(\mathbf{A}^{\text{base}},q)$ is the number of side information sets that are decodable with $q$ in $\mathbf{A}^{\text{base}}$. For a given $q$, this quantity is equal to 
 \begin{equation}
  \Set{N}(\mathbf{A}^{\text{base}},q) = \sum\limits_{i = \ell - T/k}^{\ell-1} \!\! {\ell-1 \choose i}{m - \ell \choose s - i}, \label{eq1}
 \end{equation}
 \noindent for all $q$ with $\mathbf{A}^{\text{base}}$ being non-zero, and $0$ otherwise. Since this quantity does not depend on the value of $q$, then the result follows that $\Set{D}(\mathbf{A}^{\text{base}},s) = k\ell \sum\limits_{i = \ell - T/k}^{\ell-1} \!\! {\ell-1 \choose i}{m - \ell \choose s - i}$. What remains is to prove \eqref{eq1}, which we justify as follows:
 Consider a given $q$ with a non-zero corresponding column in $A^{\text{base}}$, and let $j$ be the index of the segment to which $\mathbf{A}^{\text{base}}_q$ belongs.
 For a given side information set $\Set{S}$, let $i$ be the number of elements in $\Set{S}$ whose corresponding columns in $\mathbf{A}^{\text{base}}$ belong to $j$. 
 Then, $(q,\Set{S})$ is decodable in $\mathbf{A}^{\text{base}}$ if and only if the elements $\ell - T/k \leq i \leq \ell-1$; the lower bound is to ensure that the columns of $\mathbf{A}^\text{base}$ belonging to segment $j$ that fall outside of $\Set{S}$ are linearly independent, and the upper bound is to ensure that $q$ is not in $\Set{S}$. The number of subsets $\Set{S}$ with $i$ columns in segment $j$ is equal to ${\ell-1 \choose i}{m - \ell \choose s - \ell + 1}$. Therefore, by summing over all possible $i$ and multiplying by the number of possible requests we get the expression in \eqref{eq1}.

\section{}
\label{app::theorem_1}

For this scheme, we can have $p(\hat{\mathbf{A}}|q_1,\Set{S}_1) = 1/K$ for all $\hat{\mathbf{A}} \in \Set{A}$ for all $(q_1,\Set{S}_1) \in \Set{D}(\hat{\mathbf{A}},s_1)$, where $K$ is equal to
\begin{equation*}
K = T { s \choose \ell-1} {m - \ell \choose \underbrace{\ell \: \ell \: \cdots \: \ell}_{k-1}}^{(\text{M})},
\end{equation*}
\noindent where the last term is a multinomial coefficient.
This is because the number of column-permutations of $\hat{\mathbf{A}}^{\text{base}}$ that satisfies a given $(q_1,\Set{S}_1)$ is equal to $K$, independently of the value of $(q_1,\Set{S}_1)$. This statement can be justified as follows: for a pair to be decodable, the column of the encoding matrix corresponding to $q$ should be non-zero, and since we have $T$ segments, then there are $T$ possibilities for that column; thus the term $T$ in the expression. Next, all remaining $\ell-1$ columns of the same segment must correspond to elements in the side information set; thus the term ${ s \choose \ell-1}$. Finally, among the remaining $m - \ell$ columns, we have to choose $k-1$ segments, each of length $\ell$; thus the final multinomial term. \\

\noindent \textbf{Calculating $H(Q_1,S_1|\hat{\mathbf{A}},s_1)$:} 
Note that by using the transmission strategy described above, we satisfy the uniformity condition of Remark \ref{thm::uniformity} for \eqref{thm_prelim_joint}.
Therefore, we have $H(Q_1,S_1|\hat{\mathbf{A}},s_1) = \log |\Set{D}(\hat{\mathbf{A}},s_1)| = \log Tl {m-\ell \choose s-\ell+1}$. The last equality can be obtained by considering \eqref{scheme_dec_set} with $k = T$.\\

\noindent \textbf{Calculating $H(Q_1|\hat{\mathbf{A}},s_1)$:} 
Using the transmission strategy described above also satisfies the uniformity condition of Remark \ref{thm::uniformity} for \eqref{thm_prelim_Q}. To see this, note that 
\begin{equation*}
 p(q_1|\hat{\mathbf{A}},s_1) = \!\!\!\!\!\!\!\!\!\!\!\! \sum\limits_{\Set{S}_1 : (q_1,\Set{S}_1) \in \Set{D}(\hat{\mathbf{A}},s_1)} \!\!\!\!\!\!\!\!\!\!\!\! p(q_1,\Set{S}_1|\hat{\mathbf{A}},s_1),
\end{equation*}
\noindent where the number of elements in the summation corresponds to the number of subsets $\Set{S}_1$ that are decodable with $q_1$, which is equal to ${m - \ell \choose s-\ell+1}$ irrespective of $q_1$. Therefore, $p(q_1|\hat{\mathbf{A}},s_1)$ is uniform over all $q_1 \in \Set{D}^Q(\hat{\mathbf{A}},s_1)$.
Thus we have $H(Q_1|\hat{\mathbf{A}},s_1) = \log |D^Q(\hat{\mathbf{A}},s_1)| = \log T \ell$, where the last equality similarly holds by considering \eqref{scheme_dec_set} with $k = T$.\\

\noindent \textbf{Calculating $H(S_1|\hat{\mathbf{A}},s_1)$:} 
Using the transmission strategy above does not satisfy the uniformity condition of Remark \ref{thm::uniformity} for \eqref{thm_prelim_S}. Therefore, we now seek to quantify the achieved value of $H(S_1|\hat{\mathbf{A}},s_1)$.

Note that the used transmission strategy would yield $p(q_1,\Set{S}_1|\hat{\mathbf{A}},s_1) = 1/|\Set{D}(\hat{\mathbf{A}},s_1)|$ for all $(q_1,\Set{S}_1) \in \Set{D}(\hat{\mathbf{A}},s_1)$ and $0$ otherwise. One can then write the marginal $p(\Set{S}_1|\hat{\mathbf{A}},s_1)$ as 

\begin{equation}
 \nonumber
 p(\Set{S}_1|\hat{\mathbf{A}},s) = \sum\limits_{q_1 \in \Set{D}^Q(\hat{\mathbf{A}},s_1)} p(q_1,\Set{S}_1|\hat{\mathbf{A}},s_1) = \frac{N_{\hat{\mathbf{A}},\Set{S}_1}}{|\Set{D}(\hat{\mathbf{A}},s_1)|},
\end{equation}
\noindent where $N_{\hat{\mathbf{A}},\Set{S}_1}$ is the number of requests $q_1$ that are decodable with $\Set{S}_1$ in $\hat{\mathbf{A}}$.
Therefore, we have

\begin{align}
\label{H}
 H(\Set{S}_1| &\hat{\mathbf{A}}, s_1) = - \!\!\!\!\!\!\!\!\! \sum_{\Set{S}_1 \in \Set{D}^S(\hat{\mathbf{A}},s_1)}   \frac{N_{\hat{\mathbf{A}},\Set{S}_1}}{|\Set{D}(\hat{\mathbf{A}},s_1)|} \log  \frac{N_{\hat{\mathbf{A}},\Set{S}_1}}{|\Set{D}(\hat{\mathbf{A}},s_1)|} \nonumber \\
 &=\log |\Set{D}(\hat{\mathbf{A}},s_1)| - \frac{1}{|\Set{D}(\hat{\mathbf{A}},s_1)|} \underbrace{  \sum_{\Set{S}_1 \in \Set{D}^S(\hat{\mathbf{A}},s_1)}  \!\!\!\!\!\!\! N_{\hat{\mathbf{A}},\Set{S}_1} \log N_{\hat{\mathbf{A}},\Set{S}_1}}_{\bar{N}_t}.
\end{align}

%
Next we calculate $\bar{N}_t$. 
For a given $\Set{S}_1$, let $\ell_j, j \in [T]$ be the number of elements of $\Set{S}_1$ for which the corresponding columns in $\hat{\mathbf{A}}$ belong to segment $j$.
Then in order for a pair $(q_1,\Set{S}_1)$ to be decodable, then $\ell_j$ must be exactly equal to $\ell-1$, where $j$ corresponds to the segment to which $\hat{\mathbf{A}}_q$ belongs.

Note that $N_{\hat{\mathbf{A}},\Set{S}_1}$ only depends on the values of $\ell_j$, and therefore all subsets $\Set{S}_1$ for which $\ell_j, j \in [T]$ are the same will have the same value for $N_{\hat{\mathbf{A}},\Set{S}_1}$. Based on this fact, we can then write

\begin{align}
 \bar{N}_t &= \sum\limits_{\ell_1 = 0}^\ell \cdots \sum\limits_{\ell_T = 0}^\ell {\ell \choose \ell_1} \cdots {\ell \choose \ell_T} {m - T\ell \choose s_1 - \sum\limits_{i = 1}^T \ell_i} \left( \sum\limits_{i=1}^T \mathbb{1}_{\{\ell_i = \ell - 1 \}} \right) \log \left( \sum\limits_{i=1}^T \mathbb{1}_{\{\ell_i = \ell - 1 \}} \right)\nonumber \\
 &\stackrel{(a)}{=} \sum\limits_{x=1}^T x \log x {T \choose x} \ell^x \overbrace{\left[ \sum\limits_{ \stackrel{\ell_1 = 0}{\ell_1 \neq \ell - 1} }^\ell \!\!\! \cdots \!\!\! \sum\limits_{\stackrel{\ell_{T-x} = 0}{\ell_{T-x} \neq \ell - 1}}^\ell \!\!\! {\ell \choose \ell_1} \cdots {\ell \choose \ell_{T-x}} {m - T\ell \choose s_1 - x(\ell-1) - \sum\limits_{i = 1}^{T-x} \ell_i} \right]}^{C_{s_1,T}(T-x)} \label{app::N_bar_t}
\end{align}

\noindent where $(a)$ can be justified as follows: note that the possible values to which the term $ \sum\limits_{i=1}^T \mathbb{1}_{\{\ell_i = \ell - 1 \}}$ evaluates are $x \in [T]$ ($x = 0$ is also possible, but trivial). Moreover, it is equal to $x$ if and only if there are exactly $x$ indices from the set $\ell_{[T]}$ which are equal to $\ell-1$, while the remaining indices can take any value (except $\ell-1$). Therefore, by means of counting arguments, $\bar{N}_t$ can be expressed as \eqref{app::N_bar_t}.

Note that we can write 
\begin{align}
 &{C_{s_1,T}(T-x)} = \left[ \sum\limits_{ \stackrel{\ell_1 = 0}{\ell_1 \neq \ell - 1} }^\ell \!\!\! \cdots \!\!\! \sum\limits_{\stackrel{\ell_{T-x} = 0}{\ell_{T-x} \neq \ell - 1}}^l \!\!\! {\ell \choose \ell_1} \cdots {\ell \choose \ell_{T-x}} {m - T\ell \choose s_1 - x(\ell-1) - \sum\limits_{i = 1}^{T-x} \ell_i} \right] \nonumber \\
 &\stackrel{(b)}{=} \overbrace{\left[ \sum\limits_{ \stackrel{\ell_1 = 0}{} }^\ell \!\!\! \cdots \!\!\! \sum\limits_{\stackrel{\ell_{T-x} = 0}{}}^\ell \!\!\! {l \choose \ell_1} \cdots {\ell \choose \ell_{T-x}} {m - T\ell \choose s_1 - x(\ell-1) - \sum\limits_{i = 1}^{T-x} \ell_i} \right]}^{B_{s_1,T}(T-x)} - \nonumber \\
 & \sum\limits_{y=1}^{T-x} {T-x \choose y} \ell^y \left[ \sum\limits_{ \stackrel{\ell_1 = 0}{\ell_1 \neq \ell - 1} }^\ell \!\!\! .. \!\!\! \sum\limits_{\stackrel{\ell_{T-x-y} = 0}{\ell_{T-x-y} \neq \ell - 1}}^\ell \!\!\! {\ell \choose \ell_1} .. {\ell \choose \ell_{T-x-y}} {m - T\ell \choose s_1 - (x+y)(\ell-1) - \sum\limits_{i = 1}^{T-x-y} \ell_i} \right] 	\nonumber \\
 &= B_{s_1,T}(T-x) -  \sum\limits_{y=1}^{T-x} {T-x \choose y} \ell^y C_{s_1,T}(T-x-y) \label{app::C_rec}
\end{align}

\noindent where $(b)$ follows by adding the missing summation terms of $C_{s_1,T}(T-x)$ corresponding to $\ell_i = \ell - 1$ and - by means of counting - subtracting them. By noting that $C_{s_1,T}(0) = {m - T\ell \choose s_1 - T(\ell-1)}$, equation \eqref{app::C_rec} then defines a linear recurrence relation on $C_{s_1,T}(T-x)$ which we solve in the following lemma.

\begin{lemma}
 The solution to the linear recurrence relation in \eqref{app::C_rec} is
 \begin{equation}
  C_{s_1,T}(T-x) = \sum\limits_{v=0}^{T-x} (-1)^v \ell^v {T-x \choose v} B_{s_1,T}(T-x-v) \label{app::C_close}
 \end{equation}
 where $B_{s_1,T}(0) = {m - T\ell \choose s_1 - T(\ell-1)}$.
\end{lemma}

\begin{Pf}
 We will solve the recurrence relation using strong induction. Specifically, assume that
 \begin{equation}
  \nonumber
  C_{s_1,T}(T-x-y) = \!\!\!\! \sum\limits_{v=0}^{T-x-y} \!\!\!\! (-1)^v \ell^v {T-x-y \choose v} B_{s_1,T}(T-x-v-y)
 \end{equation}
 
 \noindent for $1 \leq y \leq T-x$. Then consider
 
 \begin{align}
  &\sum\limits_{y=1}^{T-x} {T-x \choose y} \ell^y C_{s_1,T}(T-x-y) = \nonumber \\
  & = \sum\limits_{y=1}^{T-x} \sum\limits_{v=0}^{T-x-y} \!\!\!\! (-1)^v \ell^{v+y} {T-x \choose y} {T-x-y \choose v} B_{s_1,T}(T-x-v-y) \nonumber \\
  & \stackrel{(c)}{=} \sum\limits_{k=1}^{T-x} (-1)^k \ell^k {T-x \choose k} B_{s_1,T}(T-x-k) \sum\limits_{v=0}^{k-1}  (-1)^{v-k} \frac{{T-x \choose k-v}{T-x-k+v \choose v}}{{T-x \choose k}}   \nonumber \\
  & = \sum\limits_{k=1}^{T-x} (-1)^k \ell^k {T-x \choose k} B_{s_1,T}(T-x-k) \sum\limits_{v=0}^{k-1}  (-1)^{k-v} {k \choose k-v}  \nonumber \\
  & = \sum\limits_{k=1}^{T-x} (-1)^k \ell^k {T-x \choose k} B_{s_1,T}(T-x-k) \sum\limits_{v^\prime=1}^{k}  (-1)^{v^\prime} {k \choose v^\prime}  \nonumber \\
  & = \sum\limits_{k=1}^{T-x} (-1)^k \ell^k {T-x \choose k} B_{s_1,T}(T-x-k) (\delta_{k0} - 1)  \nonumber \\
  & = -\sum\limits_{k=0}^{T-x} (-1)^k \ell^k {T-x \choose k} B_{s_1,T}(T-x-k)  +  B_{s_1,T}(T-x) \nonumber 
 \end{align}
 \noindent where $(c)$ follows by i) changing summation variables as $v + y =k$ and ii) multiplying and dividing by $(-1)^k{T-x \choose k}$, and where $\delta_{ij}$ is the Kronecher delta function. Therefore we have
 
 \begin{align}
  C_{s_1,T}(T-x) &= \sum\limits_{k=0}^{T-x} (-1)^k \ell^k {T-x \choose k} B_{s_1,T}(T-x-k) \nonumber \\
  &= B_{s_1,T}(T-x) - \sum\limits_{y=1}^{T-x} {T-x \choose y}\ell^y C_{s_1,T}(T-x-y) \nonumber
 \end{align}
\noindent satisfying \eqref{app::C_rec}, thus completing the proof.
\end{Pf}

By plugging \eqref{app::C_close} in \eqref{app::N_bar_t}, we can further simply \eqref{app::N_bar_t} as follows

\begin{align}
 \bar{N}_t &= \sum\limits_{x=1}^T x \log x {T \choose x} \ell^x \sum\limits_{v=0}^{T-x} (-1)^v \ell^v {T-x \choose v} B_{s_1,T}(T-x-v) \nonumber \\
 &= \sum\limits_{x=1}^T \sum\limits_{v=0}^{T-x} x \log x {T \choose x} {T-x \choose v} \ell^{x+v}  (-1)^v  B_{s_1,T}(T-x-v) \nonumber \\
 &= \sum\limits_{x=1}^T \sum\limits_{v=0}^{T-x} x \log x {T \choose x+v} {x+v \choose x} \ell^{x+v}  (-1)^v B_{s_1,T}(T-x-v) \nonumber \\
  &= \sum\limits_{i=1}^T \sum\limits_{x=1}^{i} x \log x {T \choose i} {i \choose x} \ell^i  (-1)^{i-x}   B_{s_1,T}(T-i) \nonumber \\
  &= \sum\limits_{i=1}^T  {T \choose i}  \ell^i B_{s_1,T}(T-i) \sum\limits_{x=1}^{i} (-1)^{i-x} {i \choose x} x \log x \nonumber \\
  &= \sum\limits_{i=1}^T  {T \choose i}  \ell^i B_{s_1,T}(T-i) \sum\limits_{x=1}^{i} (-1)^{i-x} i {i - 1 \choose x - 1} \log x \nonumber \\
  &= T \sum\limits_{i=1}^T  {T-1 \choose i-1 }  \ell^i B_{s_1,T}(T-i) \sum\limits_{x=1}^{i} (-1)^{i-x} {i - 1 \choose x - 1} \log x \label{app::N_bar_t_final}.
\end{align}

Also, we can write 

\begin{align}
 B_{s_1,T}&(T-x) = \underbrace{\sum\limits_{\ell_1 = 0}^\ell  \cdots  \sum\limits_{\ell_{T-x} = 0}^\ell \sum\limits_{y = 0}^{m-T\ell}}_{ \sum\limits_{i=1}^{T-x}l_i + y = s_1 - x(\ell-1)} {\ell \choose \ell_1} \cdots {\ell \choose \ell_{T-x}} {m - T\ell \choose y} \stackrel{(d)}{=} {m - x\ell \choose s_1 - x (\ell-1)} \label{app::vandermonde}
\end{align}

\noindent where $(d)$ follows by using Vandermonde's identity. Using \eqref{H}, \eqref{app::N_bar_t_final} and \eqref{app::vandermonde} thus proves the theorem.

\end{document}